\documentclass[12pt]{article}
=0cm \headsep =0cm

\usepackage {latexsym}
\usepackage {graphicx}

\begin{document}

\title{ {\large {\bf Anti de Sitter 5D black hole solutions with a self-interacting bulk scalar field: a potential reconstruction
approach}}   }
\author{K. Farakos \footnote{kfarakos@central.ntua.gr}, A. P. Kouretsis \footnote{a.kouretsis@yahoo.gr} and P. Pasipoularides \footnote{p.pasip@gmail.com} \\
       Department of Physics, National Technical University of
       Athens \\ Zografou Campus, 157 80 Athens, Greece}
\date{}
\maketitle
\begin{abstract}
We construct asymptotically AdS black hole solutions, with a self-interacting bulk scalar field, in the context of 5D general relativity.
As the observable universe is characterized by spatial flatness, we focus to solutions
where the horizon of the black hole, and subsequently all 3D hypersurfaces for fixed radial coordinate,
have zero spatial curvature. We examine two cases for the black hole scalar hair:
a) an exponential decaying scalar field profile and b) an inverse power scalar field profile. The scalar black hole solutions we
present in this paper, are characterized by four functions $f(r)$, $a(r)$, $\phi(r)$ and $V(\phi(r))$. Only the functions $\phi(r)$ and $a(r)$ are determined analytically, while the functions $f(r)$ and $V(\phi(r))$ are expressed semi-analytically, by integral formulas in terms of $a(r)$. We present our numerical results and study in detail the characteristic properties of our solutions. We also note that the potential we obtain has a non-convex form in agreement with the corresponding "no hair theorem" for AdS space-times.
\end{abstract}

\section{Introduction}

According to the "no hair conjecture" by R. ~Ruffini and J.~ A.~ Wheeler \cite{RW},
conventional black holes, namely vacuum solutions of the Einstein-Maxwell system,
are fully described by the parameters of mass, electric and magnetic charge and angular momentum. Parameters
such as multipole moments, which are introduced when spherical symmetry is broken, are absent after
gravitational collapse of matter sources inside the black hole event horizon.

Although black hole solutions are severely restricted in the case
of the standard Einstein-Maxwell action, new solutions can be obtained for generalized actions, which include for
example scalar or non-Abelian Gauge fields. The so called "no hair theorems" are formulated for specific
models, under certain symmetry and asymptotic behavior considerations for the metric, and set concrete restrictions to the corresponding vacuum solutions.

A neutral scalar field with a self-interaction potential term $V(\phi)$ is served as a first example of a study beyond the
Einstein-Maxwell action. The corresponding "no hair theorem" has been formulated long ago by Bekenstein in Ref. \cite{Bekenstein:1971hc}.
Accordingly, there is no asymptotically flat black hole solution with a nontrivial continuous scalar "hair" for a convex potential
$V(\phi)$ ($V'(\phi)\geq0$). However, non-trivial scalar black hole solutions can be obtain, if we relax some of the above conditions
of the "no hair theorem".  Recent (analytical or numerical)
asymptotically flat solutions in four dimensions with "scalar hair" are presented in Refs. \cite{Bechmann:1995sa,Dennhardt:1996cz,Bronnikov:2001ah,Nucamendi:1995ex,Hertog:2006rr,Kuriakose:2008dj},
while asymptotically AdS solutions can be found for example in Refs. \cite{  Martinez:2004nb,Dotti:2007cp, Torii:2001pg,Winstanley:2002jt,Winstanley:2005fu,Zeng:2009fp }. The issue of the stability, against
linear perturbation around the scalar field, have been also examined, see Refs. \cite{Dennhardt:1996cz,Torii:2001pg,Winstanley:2002jt,Winstanley:2005fu,Harper:2003wt,Hertog:2004bb}.

We would like to emphasize, that the scalar black hole solutions, which are presented in the above mentioned references,
do not introduce new quantities which could characterize the black hole, beyond the standard ones of
mass, electric (or magnetic) charge and angular momentum. Hence, the meaning of the "no hair conjecture" remains, see also the relative discussions in Refs. \cite{Mavromatos:1995fc, Weinberg:2001gc}.  Note that in Ref. \cite{Bekenstein:1974sf} an analytical solution with a scalar field conformally coupled to gravity which introduces a new quantity (in particular a scalar charge) is presented, but the scalar field blows up on the event horizon of the black hole, see also the comments in Ref. \cite{Kuriakose:2008dj}.

In this paper we study 5D asymptotically AdS black hole solutions with scalar "hair" in a semi-analytical way. We focus to solutions where all 3D hypersurfaces for fixed radial coordinate $r$ have zero spatial curvature, in contrast with the usual case where the horizon of the black hole is characterized by spherical symmetry (or with positive spatial curvature). In the present approach the self-interaction potential $V(\phi)$ of the
scalar field is assumed to be an undetermined function, and there is a freedom in the choice of the scalar field profile $\phi(r)$. Note that the function $a(r)$ and $f(r)$ which appear in the black hole metric, as well as the potential $V(\phi)$, depend on the specific choice of the scalar field. We have examined two cases for the scalar field: (a) an exponential decaying profile and (b) an inverse power profile, of Eqs. (\ref{exp}) and (\ref{coulomp}) below correspondingly.  Note that these profiles ((a) and (b) above)  have been studied previously by Lechtenfeld and co-authors in Refs. \cite{Bechmann:1995sa,Dennhardt:1996cz}, for 4D black hole solutions in asymptotically flat space-time, with an analogous methodology that we follow in this paper. In the solutions we present the functions $\phi(r)$ and $a(r)$ are determined analytically, while the functions $f(r)$ and $V(\phi(r))$ are expressed semi-analytically by integral formulas in terms of $a(r)$. In the main part of this work we present our results in figures and the characteristic properties of our solutions are discussed. We see that the reconstructed potential $V(\phi)$, for the specific choices of the scalar field (a) and (b), has a non-convex form in agreement with the corresponding "no hair theorem" for $AdS_5$ space-times which is presented in section 4 of this work.

The interest for five dimensional AdS black holes is motivated by extra dimensional theories, and mainly by the so called brane world models \cite{antoniadis,ArkaniHamed:1998rs, Antoniadis:1998ig,Randall:1999ee,Randall:1999vf} which promise a resolution for the  hierarchy problem. In the case we examine our world is assumed to be trapped in a hypersurface of fixed radius $r$ in the background of a five dimensional AdS black hole vacuum \cite{Csaki:2000dm}. This also explains why we have focused to solutions for which spatial 3D sections are characterized by zero curvature, in agreement with the current astrophysical phenomenology. Note, that in contrast with the standard vacuum of Randall-Sundrum \cite{Randall:1999ee,Randall:1999vf}, the five dimensional AdS black hole vacuum does not preserve 4D Lorentz invariance on the brane, which may have interesting phenomenological implications, see for example Refs. \cite{Csaki:2000dm,Farakos:2008rv}. Note also, that in the framework of AdS/CFT correspondence, an $AdS_5$ black hole background is of particular interest as it can trigger a thermal conformal field theory on the $AdS_5$ boundary (brane), see Ref. \cite{Gubser:1999vj}.

\section{5D AdS black holes with a self-interacting bulk scalar field}

We consider the following 5D action
 \begin{eqnarray}
 S=\int d^{5}x\sqrt{|g|}\left(\frac{1}{2 \kappa_5 }R-\frac{1}{2}g^{\mu\nu}\nabla_{\mu}\phi\nabla_{\nu}\phi-V(\phi)\right), \quad \mu,\nu=0,1,2,3,5
 \end{eqnarray}
for general relativity, with a bulk self-interacting scalar field with a potential $V(\phi)$, where $\kappa_5=8 \pi G_5$ ($G_5$ is
the 5D Newton constant). Note that the extra dimension is parameterized by the
coordinate $x^{5}=r$ (radius of the black hole), while the other coordinates $x^{0},\: x^{1},\: x^{2} ,\:x^{3}$ corresponds to the usual 4D space-time. In addition we have assumed that a negative cosmological constant $\Lambda$ is incorporated in the potential of the scalar field, according to the equation $ \Lambda=V(0)$ ($V(0)<0$).

 The Einstein equations for the above action read:
 \begin{eqnarray}
 R_{\mu\nu}-\frac{1}{2}g_{\mu\nu}R=\kappa_5 T^{(\phi)}_{\mu\nu}\label{field1}
 \end{eqnarray}
and the energy momentum tensor $T^{(\phi)}_{\mu\nu}$ for the bulk scalar field is
 \begin{eqnarray}
 T^{(\phi)}_{\mu\nu}=\nabla_{\mu}\phi\nabla_{\nu}\phi-
 g_{\mu\nu}[\frac{1}{2}g^{\rho\sigma}\nabla_{\rho}\phi\nabla_{\sigma}\phi+V(\phi)]\label{energymomentum}
 \end{eqnarray}
 If we use Eqs. (\ref{field1}) and (\ref{energymomentum}) we obtain the equivalent equation:
 \begin{eqnarray}
 R_{\mu\nu}-\kappa_{5}\left(\partial_\mu\phi \partial_\nu\phi-\frac{2}{3}g_{\mu\nu}V(\phi)\right)=0 \label{einstein1}
 \end{eqnarray}
Now for the metric of the black hole solution we make the following ansatz
 \begin{eqnarray}
 ds^{2}=-f(r)dt^{2}+f^{-1}(r)dr^{2}+a^{2}(r)d\textbf{x}^2 \label{metricBH}
 \end{eqnarray}
where $d\textbf{x}^2$ is the metric of the spatial 3-section, which in our case are assumed to have zero curvature, in agreement with the current astrophysical phenomenology, pointing toward spatial flatness of the observable universe.

In the case of the metric of Eq. (\ref{metricBH}), if we use Eq. (\ref{einstein1}) we find the
following three independent differential equations
 \begin{eqnarray}
 f''(r)+3\frac{a'(r)}{a(r)}f'(r)+\frac{4}{3}V(\phi)=0\label{first}
 \end{eqnarray}
\begin{eqnarray}
\frac{a'(r)}{a(r)}f'(r)+\left(2\frac{\left(a'(r)\right)^2}{a^{2}(r)}+\frac{a''(r)}{a(r)}\right)f(r)+\frac{2}{3}V(\phi)=0\label{second}
\end{eqnarray}
\begin{eqnarray}
f''(r)+3\frac{a'(r)}{a(r)}f'(r)+6\left(\frac{a''(r)}{a(r)}+\frac{1}{3}(\phi'(r))^{2}\right)f(r)+\frac{4}{3}V(\phi)=0\label{third}
\end{eqnarray}
All the quantities, in the above equations, have been rendered dimensionless via the redefinitions
$\sqrt{\kappa_5} \phi\rightarrow \phi$, $\kappa_5 \ell^{-2} V \rightarrow V$ and $r/l \rightarrow r$,
where the $AdS_5$ radius $l$ in the above rescaling is defined as $l=\sqrt{-6/(\kappa_5 \Lambda)}$.

If we eliminate the potential $V(\phi)$ from the above equations we obtain:
\begin{eqnarray}
a''(r)+\frac{1}{3}(\phi'(r))^2 a(r)=0\label{adiff}
\end{eqnarray}
\begin{eqnarray}
f''(r)+f'(r)\frac{a'(r)}{a(r)}-\left(4\frac{(a'(r))^2}{a^{2}(r)}+2\frac{a''(r)}{a(r)}\right)f(r)=0\label{fdiff}
\end{eqnarray}
where the potential can be determined from Eq. (\ref{first}) if the functions $a(r)$ and $f(r)$ are known.

We observe that in the above two differential equations (\ref{adiff}) and (\ref{fdiff}) we have three unknown functions, hence we have the freedom to choose one of them, for example the scalar field $\phi(r)$, and the other two functions $a(r)$ and $f(r)$ can be determined subsequently.
We will consider continuous scalar field deformations which are localized in a small region of r, while for large values of r they tend rapidly to zero. Mainly, we aim to study the following two cases for the scalar field:
a) an exponential profile and b) an inverse power profile, which are given by the following equations
\begin{eqnarray}
&&\phi_1(r)=\phi_0 e^{-\frac{r}{d}}\label{exp}\\
&&\phi_2(r)=\frac{q}{r^n} \label{coulomp}
\end{eqnarray}
As we will see in the next section the above choices of Eqs. (\ref{exp}) and (\ref{coulomp}) can lead to analytic solutions for the function $a(r)$ by solving Eq. (\ref{adiff}). However, the other functions $f(r)$ and $V(\phi)$ can be determined semi-analytically. Note that these profiles have been studied previously by the authors of Refs. \cite{Bechmann:1995sa,Dennhardt:1996cz}
in a similar problem for 4D black hole solutions with a self-interacting phantom
scalar field.

Now, if the function $a(r)$ has been obtained for a specific choice of the scalar field $\phi(r)$, the
function $f(r)$ can be determined by Eq. (\ref{fdiff}), for which the general solution can be expressed as a
linear combination of two independent solutions:
\begin{equation}
f(r)=C_{1}a^2(r)+C_{2}a^2(r)\int_r^{+\infty}\frac{dr'}{a^{5}(r')}\label{fr}
\end{equation}
In order to fix the constants of integration $C_{1}$ and $C_{2}$ we assume that for large r the solution approaches the well known $AdS_{5}$ Schwarzschild black hole solution. For the asymptotic behavior, see for example \cite{Csaki:2000dm, Farakos:2008rv}, and references there in. Thus we will look for solutions with the following asymptotic behavior:
 \begin{eqnarray}
&& a(r)\rightarrow r\label{asya}\\
&& f(r)\rightarrow r^2-\frac{ \mu}{r^2} \label{asyfr}
\end{eqnarray}
where $\mu$ is a dimensionless constant of integration \footnote{ The mass of the black hole, without the scalar "hair", is $m=(8 \pi G_5)^{-1}\ell^2 \mu$.}.
Note, that the previous consideration for the asymptotic behavior of the black hole solution is reasonable, because the scalar field vanishes in the infinity and the scalar field potential $V(\phi)$ approaches a negative non zero value $\Lambda=V(0)$ which corresponds to the 5D cosmological constant. If we compare Eq. (\ref{fr}) and Eq. (\ref{asyfr}) in the limit of large $r$ we find:
 \begin{equation}
 C_{1}=1,\quad
C_{2}=-4\mu
\end{equation}
hence we obtain the formula
 \begin{equation}
f(r)=a^2(r)\left[1-\int_r^{+\infty}\frac{4\mu}{a^{5}(r')}dr'\right]\label{fr2}
\end{equation}
Note that the integration over $r'$ is restricted in the range $r'>r_{s}$, where $r_{s}$ is the largest zero of the function $a(r)$ that represents
the physical singularity of the black hole. The horizon of the black hole $r_h$ can be determined via the equation $f(r_h)=0$, or equivalently by the
equation
 \begin{equation}
\int_{r_{h}}^{+\infty}\frac{1}{a^{5}(r')}dr'=\frac{1}{4\mu}\label{fr3}
\end{equation}
which always has a unique positive solution.

By replacing Eq. (\ref{fr2}) into Eq. (\ref{first}) the scalar field potential $V(\phi(r))$ can be expressed in terms of the function $a(r)$, according to the equation
 \begin{equation}
 V(\phi(r))=-6\mu\frac{a'(r)}{a^4(r)}+\left(4\mu\:\int^{+\infty}_{r}\frac{1}{a^5(r)}\:dr-1\right)\left(6(a'(r))^2+
 \frac{3}{2}a''(r)a(r)\right)\label{Vr}
 \end{equation}

 \begin{figure}[h]
\begin{center}
\includegraphics[width=0.75 \textwidth, angle=0]{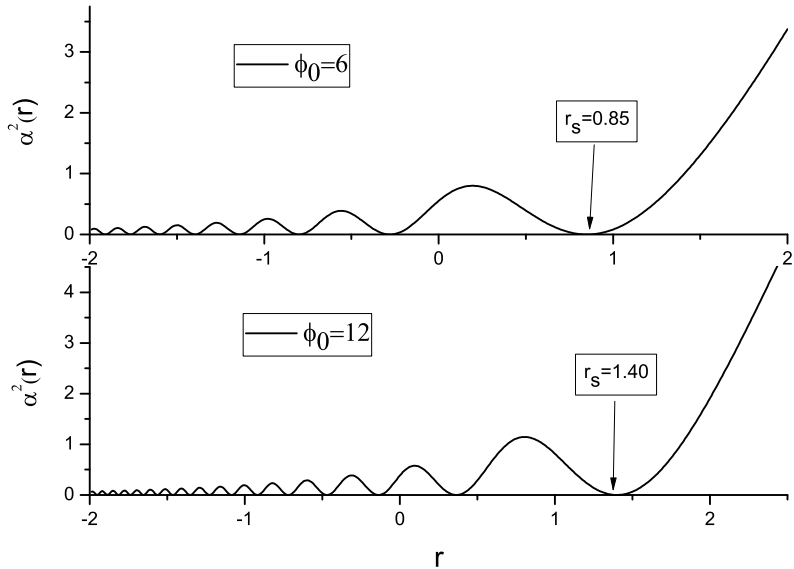}
\end{center}
\caption{\small The factor $a^2(r)$ as a function of $r$ for $d=1$ and $\phi_{0}=6,12$. We
observe that the position of the physical singularity $r_s$ becomes larger when $\phi_0$ increases.}\label{1}
\end{figure}

\begin{figure}[h]
\begin{center}
\includegraphics[width=0.75 \textwidth, angle=0]{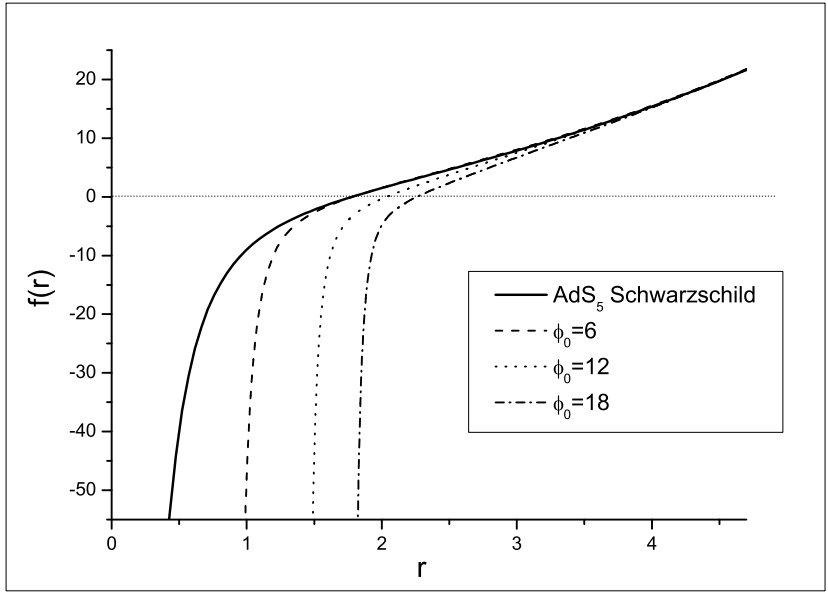}
\end{center}
\caption{\small The factor $f(r)$ as a function of $r$ for $\mu=10$,  $d=1$ and $\phi_{0}=6,12,18$. We
observe that for $r\rightarrow +\infty$ we recover the well known $AdS_5$ Schwarzschild black hole solution.}\label{2}
\end{figure}

\begin{figure}[h]
\begin{center}
\includegraphics[width=0.75 \textwidth, angle=0]{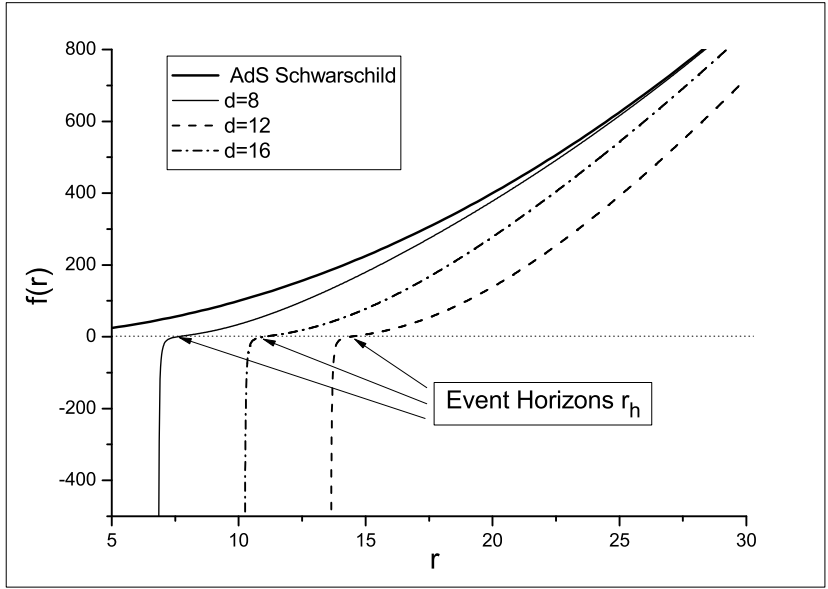}
\end{center}
\caption{\small The factor $f(r)$ as a function of $r$ for $\mu=10$,  $\phi=6$ and $d=8,12,16$. We
observe that when $d>>1$, outside the event horizon, the scalar black hole differs significantly from the $AdS_5$ Schwarzschild black hole (bold
line in the figure). Although it is not displayed in the figure, we have checked that for enough large $r$ we recover the well known $AdS_5$ Schwarzschild black hole solution, even for d=12 and d=16.}\label{25}
\end{figure}

\begin{figure}
\begin{center}
\includegraphics[width=0.75 \textwidth, angle=0]{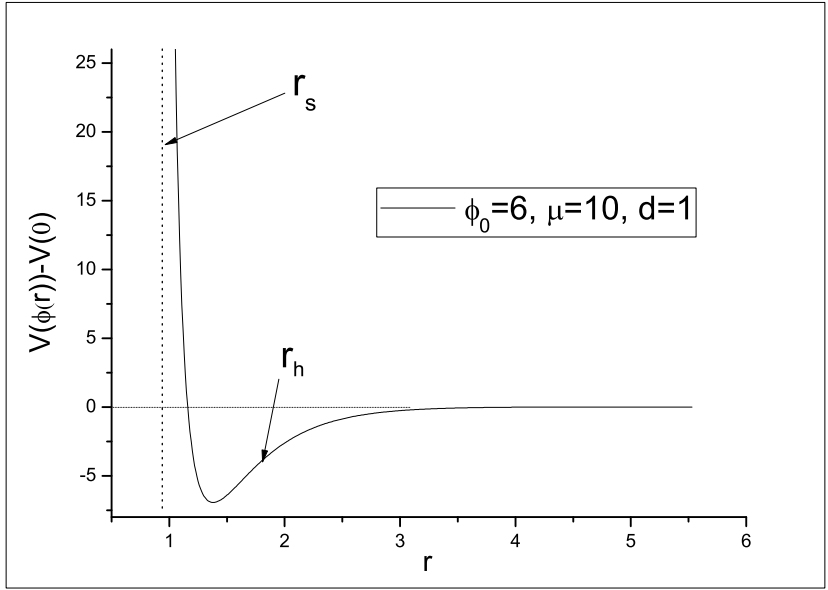}
\end{center}
\caption{\small The potential $V(\phi(r))$ as a function of $r$ for $\mu=10$,  $d=1$ and $\phi_{0}=6$.
It blows up for $r\rightarrow r_{s}$, while for large $r$ it tends to a constant value equal
to 5D cosmological constant. We also see that minimum value of the potential is inside the black hole horizon $r_h$. }\label{3}
\end{figure}

\begin{figure}[h]
\begin{center}
\includegraphics[width=0.75 \textwidth, angle=0]{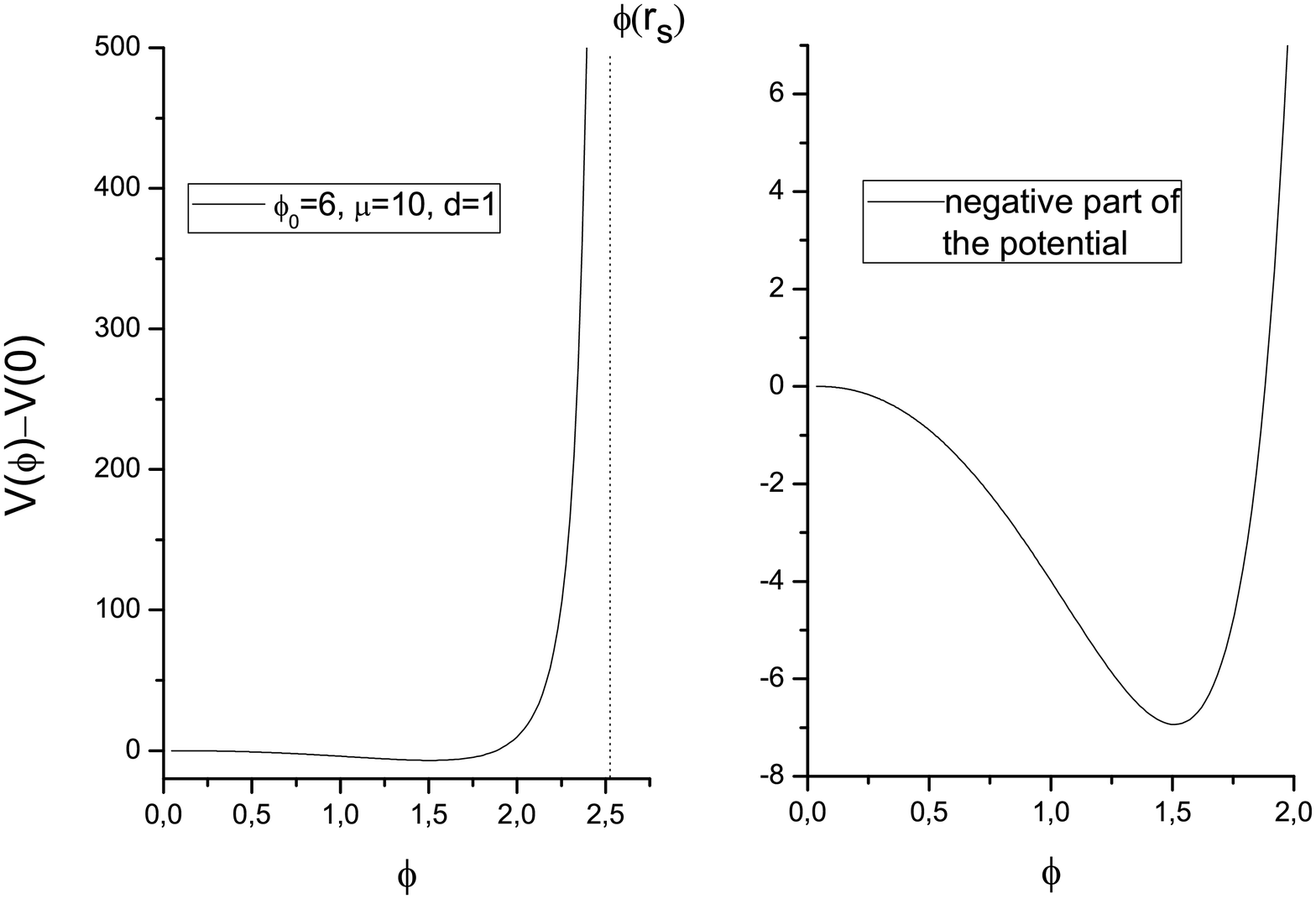}
\end{center}
\caption{\small The potential $V(\phi)-V(0)$ as a function of $\phi$ for $\mu=10$,  $d=1$ and $\phi_{0}=6$.
In the left panel we observe that the potential blows up at the point where the scalar field becomes equal to  $\phi(r_{s})$.
In the right panel we have plotted in detail the negative part of the potential ($V(\phi)-V(0)$) and we see that it possesses a local minimum in the range $\phi(r_h)<\phi<\phi(r_s)$.}\label{4}
\end{figure}

\section{Numerical Analysis}

We will study two classes of solutions for the profiles of Eqs. (\ref{exp}) and (\ref{coulomp}) above. The warp factor
$a(r)$ can be determined analytically by the differential equation (\ref{adiff}),  while the black hole factor $f(r)$
and the potential $\hat{V}(r)=V(\phi(r))$ will be computed by the integral formulas of Eqs. (\ref{fr2}) and (\ref{Vr}). The
potential $V(\phi)$ as a function of the scalar field $\phi$ can be determined by the replacement $r=r^{-1}(\phi)$ in the function $\hat{V}(r)$.

\subsection{The exponential profile of Eq. (\ref{exp})}

In the case of the exponential profile $\phi_1(r)=\phi_{0}e^{-\frac{r}{d}}$, from Eq. (\ref{adiff}) we get the analytic solution
\begin{eqnarray}
a(r)=\tilde{C}_{1}J_{0}\left(\frac{\phi_{0}e^{-\frac{r}{d}}}{\sqrt{3}}\right)+\tilde{C}_{2}Y_{0}\left(\frac{\phi_{0}e^{-\frac{r}{d}}}{\sqrt{3}}\right)\label{scalefactor}
\end{eqnarray}
where the constants of integration $\tilde{C}_1$ and $\tilde{C}_2$ are fixed if we take into account the asymptotic behavior $a(r)$
in the large $r$ limit, see Eq. (\ref{asya}) above. If we expand the Bessel functions $J_0(x)$, $Y_0(x)$  for small argument $x$ (large $r$) we find
\begin{equation}
\tilde{C}_{1}=d\left[ln(\frac{\phi_{0}}{2\sqrt{3}})+\gamma\right], \quad \tilde{C}_{2}=-\frac{\pi d}{2} \label{coe1}
\end{equation}
As we see in Fig. \ref{1} the function $a^2(r)$ is an oscillating function with an infinite number of zeroes. Note that $r$ can be
negative, as it is just a coordinate and not the "real" radius of the black hole, which is represented by the function $a(r)$. The next step is to determine the largest zero $r_s$ of $a^2(r)$ ($a(r_s)=0$) which is the physical singularity of the black hole. It is reasonable to ignore completely the part of $a(r)$ for $r\leq r_s$ which has not a physical meaning, and to keep only the region for $r>r_s$ which corresponds to
the 5D AdS black hole solution. Also in Fig. \ref{1} we see that, for fixed $d$, the position $r_s$ of the singularity becomes larger when $\phi_0$ increases.

As the physical space of the parameter $r$ is restricted for $r>r_s$ the maximum value of the scalar field $\phi(r)$ is not $\phi_0$,
the maximum value for the scalar field $\phi_{max}$ is given by the equation:
\begin{equation}
\phi_{max}=\phi_{0}\: e^{-\frac{r_s}{d}}
\end{equation}
which is exponentially suppressed by the factor $e^{-\frac{r_s}{d}}$. We have checked numerically that $\phi_{max}$
is an increasing function but depends very weakly on $\phi_0$. For example for $\phi_0=50$ we obtain $\phi_{max}=2.68$, and
for $\phi_0=1000$ we obtain $\phi_{max}=3.74$. This can be explained if we take into account that the coefficient $\tilde{C}_1$
in Eq. (\ref{scalefactor}) is logarithmically dependent on $\phi_0$.

In Fig. \ref{2} we have plotted the function $f(r)$ for several values of $\phi_0$ assuming that the value of $d$ is fixed. Note that $f(r)$
has a singularity and an event horizon as a conventional black hole solution.
In this figure we see that outside the horizon, $f(r)$ is weakly dependent on the parameter $\phi_0$. Such a behavior is expected, because the parameter which has impact on $f(r)$ is the maximum value of the scalar $\phi_{max}$ and not the parameter $\phi_0$. Note that $\phi_{max}$ weakly dependes on $\phi_0$ and remains comparatively small even for very large values of $\phi_0$, as it is mentioned in the previous paragraph.

On the other hand, in Fig. \ref{25} we see that the function $f(r)$ is strongly dependent on $d$, for fixed $\phi_0$. For large $d$ ($d$ is an estimate for the size of the scalar field), in particular for $d>>\sqrt[4]{\mu}$, we observe that outside the horizon, the black hole solution becomes significantly different from the corresponding $AdS_5$ black hole solution without the scalar field (bold line in the figure). Note that for $d\sim \sqrt[4]{\mu}$, as we see in Fig. \ref{2}, $f(r)$ tends rapidly to its asymptotic behavior $r^2-\mu/r^2$ for $r>r_h$ ($r_h$ is the position of the event horizon of the black hole with the scalar field).

The potential $V(\phi(r))$, if we subtract the contribution from the cosmological constant $\Lambda=V(0)$, has been plotted
as a function of the radius coordinate $r$ in Fig. \ref{3}. It has two characteristic features: a) it blows up near the singularity of the black hole $r_s$, and b) it tends to a constant value equal to the 5D cosmological constant $\Lambda$ when $r$ tends to infinity. Now we can obtain the potential as
a function of $\phi$ by constructing the parametric plot $(\phi(r),V(\phi(r))-V(0))$, as it is presented in Fig. \ref{4}. In the left panel we see
that the potential becomes infinitely large near $\phi(r_s)$, hence the scalar field is restricted in the region $0\leq \phi \leq \phi(r_s)$. In the right panel we see that the potential has minimum and the difference $V(\phi(r))-V(0)$ becomes negative in a large region near the axes origin, namely the potential has a non-convex form in agreement with Bekenstein 's "no hair theorem".

Also we have performed numerical computation by covering other ranges of the free parameters of the model, and we have checked that the above general features, which are exhibited in figures \ref{1}, \ref{2}, \ref{3} and \ref{4},  are preserved.

\begin{figure}[h]
\begin{center}
\includegraphics[width=0.75 \textwidth, angle=0]{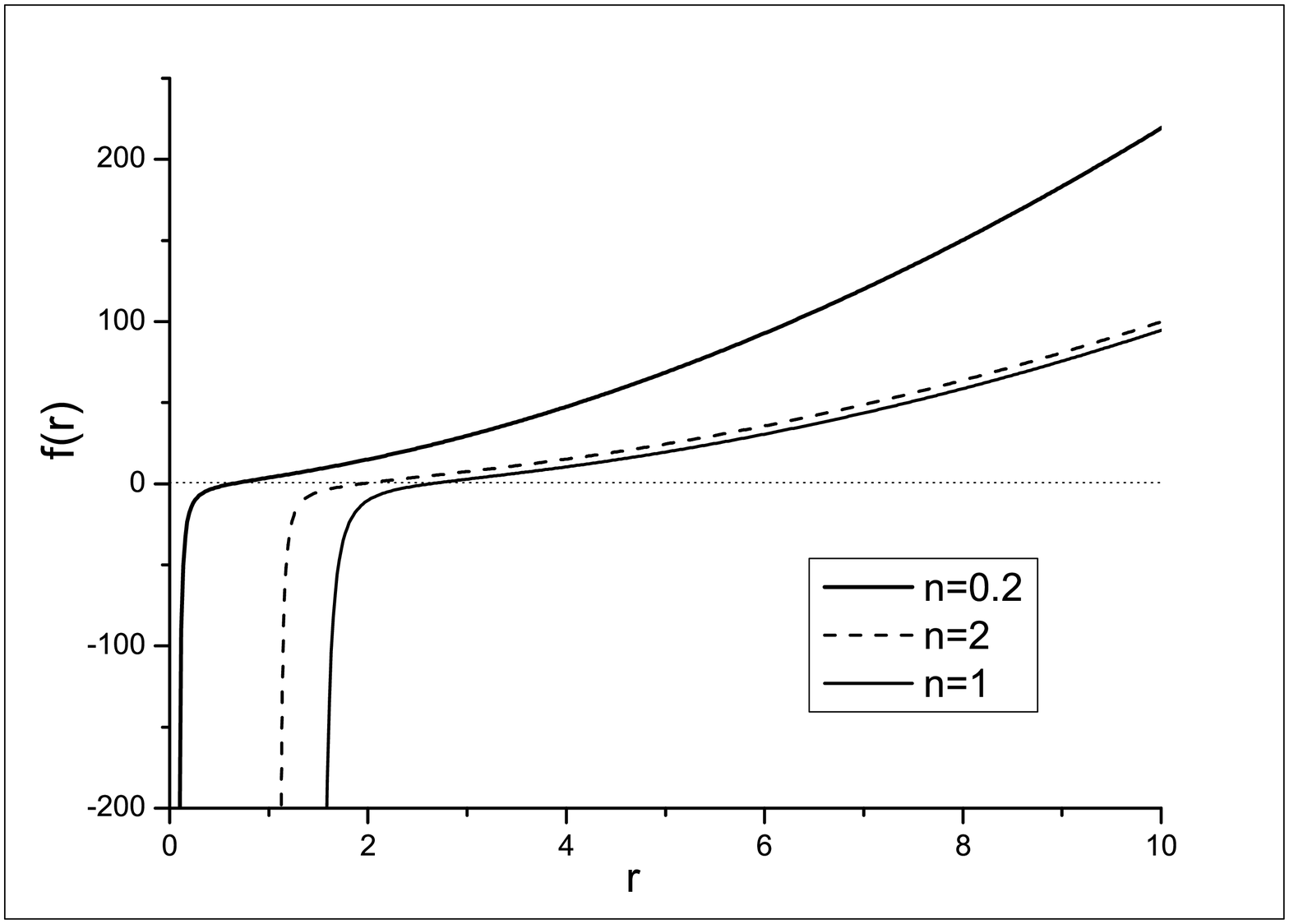}
\end{center}
\caption{\small The function f(r) for the inverse power type scalar field, for $n=0.2,~1,~2$, $q=4$ and $\mu=10$. We
observe that for $n=0.2$ and $n=1$ the asymptotic behavior of the black hole is not of the standard form $r^2-\mu/r^2$.}\label{5}
\end{figure}

\begin{figure}[h]
\begin{center}
\includegraphics[width=0.75 \textwidth, angle=0]{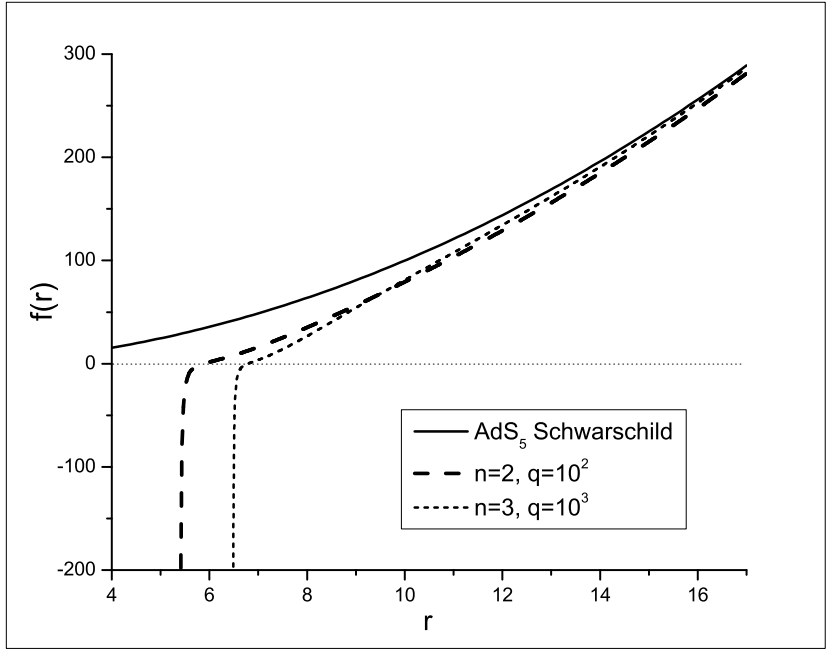}
\end{center}
\caption{\small The function f(r) for the inverse power type scalar field for $n=2$, $q=10^2$ and $n=3$, $q=10^3$ and $\mu=10$. We
see that for $n=2,3$ the asymptotic behavior of the black hole is of the standard form $r^2-\mu/r^2$, as it is expected by
the asymptotic formula of Eq. (\ref{coulompasympt}).}\label{6}
\end{figure}

\subsection{The inverse power profile of Eq. (\ref{coulomp})}

For the inverse power profile $\phi_2(r)=q/r^n$
the differential equation (\ref{adiff}) can be solved analytically with solution
\begin{equation}
a(r)=\tilde{C}_{3}\sqrt{r}\:J_{-\frac{1}{2n}}\left(\frac{q}{\sqrt{3}\: r^n}\right)
+\tilde{C}_{4}\sqrt{r}\:Y_{-\frac{1}{2n}}\left(\frac{q}{\sqrt{3}\: r^n}\right)\label{ancoulomb}
\end{equation}
A comparison with the asymptotic behavior of Eq. (\ref{asya}) implies:
\begin{equation}
\tilde{C}_{4}=0, \quad \tilde{C}_{3}=\Gamma\left(1-\frac{1}{2n}\right)\left(\frac{q}{2 \sqrt{3}}\right)^{\frac{1}{2n}}
\end{equation}
where we have assumed that $n\neq 1/2(k+1), (k=0,1,2..)$.
Note that for $n=1$ the function $a(r)$ takes the simple form
\begin{equation}
a(r)=r \:\cos\left(\frac{q}{\sqrt{3}\:r}\right)
\end{equation}
However, as we will see subsequently the value of $n$ is restricted according to the relation $n\geq 2$.

The functions $a(r)$, $f(r)$ and $V(\phi)$ for the scalar field with the inverse power profile, have similar
characteristic features with those of the exponential profile scalar field that was examined in the previous subsection. For this reason
we have restricted the number of figures in this section.
However, in the case of the inverse power profile an additional analysis is required in order to satisfy the asymptotic behavior of $a(r)$ and $f(r)$, as it is given by Eqs. (\ref{asya}) and (\ref{asyfr}) above, or we can write
\begin{equation}
a(r)\rightarrow r, \quad  f(r)\rightarrow r^2-\frac{ \mu}{r^2}+O\left(\frac{1}{r^c}\right),~~~c\geq 2 \label{asyfr1}
\end{equation}
Note that the next to leading term $r^2$, in Eq. (\ref{asyfr1}) for the function $f(r)$, is assumed to tend to zero like $1/r^2$.
Now if we take for granted an asymptotic behavior $q/r^n$ for the scalar field we can satisfy Eq. (\ref{asyfr1}) only if we set concrete restrictions to the power $n$.
For large $r$, the function $a(r)$ can be written as:
\begin{equation}
a(r)\simeq r+\delta h(r) \label{aprorxa}
\end{equation}
where $\delta h(r)$ is assumed to be a small correction if it is compared with $r$ ($\delta h(r)<<r$).
If we replace $a(r)$ in Eq. (\ref{adiff}), in the case of the inverse power profile scalar field $\phi(r)=q/r^n$, we
obtain that the correction $\delta h(r)$ is given by the equation:
\begin{equation}
\delta h(r)=-\frac{c_n}{r^{2n-1}}, ~~~~ c_{n}=\frac{n~q^2 }{6(2n-1)}
\end{equation}
In the special case where $n=1/2$ we obtain that
\begin{equation}
\delta h(r)=\frac{q^2}{12} \ln(r)
\end{equation}
For $n>1/2$ we have $2n-1>0$, and then $\delta h\rightarrow 0$ for $r\rightarrow +\infty$. This is a first restriction for the
asymptotic behavior of the scalar field. However from $f(r)$ we will set a stronger restriction as we will see bellow.

By replacing $a(r)\simeq r+\delta h(r)$ in Eq. (\ref{fr2}), if we keep only linear terms in $\delta h(r)$, we obtain the corresponding
asymptotic formula for $f(r)$:
\begin{equation}
f(r)\simeq r^2-\frac{\mu}{r^2}-\frac{2 c_n}{r^{2 n-2}}+... \label{coulompasympt}
\end{equation}
where we have neglect the terms which vanish faster than $1/r^{2n-2}$. In
order to guaranty the behavior of the  $AdS_{5}$ Schwarzschild black hole $r^2-\mu/r^2$, in the large $r$ limit,
we have to impose the stronger restriction $n>2$, hence the additional term $1/r^{2n-2}$ in Eq. (\ref{coulompasympt}) vanish
faster than $1/r^2$. Note that in the case where $n=2$ the asymptotic behavior is of the
form $r^2-\mu_{eff}/r^2$  where we have defined an effective mass $\mu_{eff}=\mu+2 c_n$.

In Fig. \ref{5} we have plotted the function $f(r)$, in the case of the inverse power scalar field profile, for $n=0.2,1,2$.
We present this figure in order to check the asymptotic formula of Eq. (\ref{coulompasympt}) above. We see that for $n=0.2$ the asymptotic behavior is not of the form $r^2-\mu/r^2$, but we have an asymptotic behavior $r^2+2 |c_{0.2}|r^{1.6}$ which
is in agrement with the formula of Eq. (\ref{coulompasympt}). In particular, the deviation between the curves with $n=0.2$ and $n=2$ in Fig. \ref{3} is due exactly the term $r^{1.6}$ in previously mentioned asymptotic behavior. For $n=1$ we observe an asymptotic behavior of the form $r^2-2 c_{1}$. This behavior can be confirmed in Fig. \ref{5} if we compare the $n=1$ and $n=2$ curves for $f(r)$ which have a constant difference. Note, that the curve with $n=2$ exhibits an asymptotic behavior of the form $r^2-\mu/r^2$ ($AdS_5$ Schwarzschild black hole) as we see in Fig. \ref{6}. In the same figure we see also that for $n=3$ we have a similar asymptotic behavior.

Now by replacing $a(r)=r+\delta h(r)$ in Eq. (\ref{Vr}) we can obtain the asymptotic formula for the potential $V(\phi)$ when $\phi \rightarrow 0$,
\begin{equation}
V(\phi)\simeq V(0)+\frac{1}{2}m^2_n \phi^2+...~~,~~~ m^2_n=n\; (n-4) \label{vsmall}
\end{equation}
We have neglected higher order terms of the form $\phi^c$ with $c>2$. Note that for $n>4$ the mass term in Eq. (\ref{vsmall}) becomes
positive, however the non-convex nature of the potential arises for larger values of $\phi$, as we have checked numerically. For $n=4$ the mass term becomes zero and the higher order terms become significant, but even in this case we obtained numerically
that the potential is non-convex.

Now it is worth to compare with the exponential profile scalar field of Eq. (\ref{exp}), for which we find
\begin{equation}
a(r)\simeq r+\delta h(r), ~~~ \delta h(r)=\frac{\phi_{0}^{2}}{12} (d+r) e^{-\frac{2 r}{d}}
\end{equation}
while the corresponding asymptotic formula for $f(r)$ is:
\begin{equation}
f(r)\simeq r^2-\frac{\mu}{r^2}+\frac{\phi_{0}^{2}}{6} (d~r+r^2) e^{-\frac{2 r}{d}}+...
\end{equation}
For $\phi(r)=\phi_{0}e^{-r/d}$ we see that the functions $a(r)$ and $f(r)$ approach their asymptotic behavior in an exponentially fast way independently from the values of the parameters $\phi_{0}$ and $d$, in contrast with the case of
the inverse power profile $1/r^n$ for which we have to set restrictions to the parameter $n$ ($n\geq 2$). The potential $V(\phi)$, in the case of the exponential profile scalar field, for $\phi \rightarrow 0$ is:
\begin{equation}
V(\phi)\simeq V(0)+\frac{1}{2} \left[\ln\left(\frac{\phi}{\phi_0}\right)\right]^2 \phi^2+... \label{vexp}
\end{equation}
where we have neglected terms of the form $\ln(\phi) \phi^2$ and $\phi^2$ as they tend to zero faster than $[\ln(\phi)]^2 \phi^2$.
Note that the above equation for small $\phi$ implies a positive derivative $V'(\phi)$, however it becomes negative for slightly larger values of
$\phi$, with $r>r_h$, as we have confirmed by numerical calculations. In left panel of Fig. \ref{4} we have plotted the negative part of $V(\phi)-V(0)$ for small $\phi$. Note that the region where $V'(\phi)$ is positive, due to the term $[\ln(\phi)]^2 \phi^2$ in Eq. (\ref{vexp}), is very closely to axes origin and it is not visible in this figure.

\section{The non-convex nature of the scalar field potential}
The equation for the scalar field
\begin{equation}
\frac{1}{\sqrt{g}}\partial_{\mu}\left(\sqrt{g} g^{\mu\nu}\partial_\nu \phi\right)
 -V'(\phi)=0
\end{equation}
can be used in order to demonstrate a "no hair theorem" like that of Bekenstein in the
special case of 5D AdS space-time we examine (see also \cite{Bechmann:1995sa} in the case of Minkowski space time). If we multiply by $\phi$ and integrate by parts, from $r_h$ (horizon of the black hole) to infinity, we
obtain
\begin{equation}
\left(g^{rr} \sqrt{g} \phi \partial_{r} \phi \right)_{r_{h}}^{+\infty}-\int_{r_h}^{+\infty} dr g^{rr} \sqrt{g}\phi (\partial_r\phi)^{2}=\int_{r_h}^{+\infty} dr \sqrt{g} \phi V'(\phi) \label{intpart}
\end{equation}
The boundary term vanishes \footnote{For $n=2$ the scalar field behaves like $1/r^2$ for large $r$. In this case the boundary term is
nonzero but it possesses a finite negative value. In particular we find that $\int_{r_h}^{+\infty} dr \sqrt{g} \phi V'(\phi)=-\int_{r_h}^{+\infty} dr g^{rr} \sqrt{g}\phi (\partial_r\phi)^{2}-2$, or equivalently that $V'(\phi)<0$. This means that the potential is non-convex even in the limiting case where $n=2$}. Because in the previous paragraph, in order to satisfy the asymptotic behavior of Eq. (\ref{asyfr1}) for the function $f(r)$, we impose the restriction $\phi(r) \simeq 1/r^n$ with $n>2$, for $r\rightarrow +\infty$  (namely $\phi(r)$ tends to zero faster than $1/r^2$).
In addition the scalar field $\phi(r)$ and the first derivative of the scalar field $\partial_{r}\phi(r)$ are regular on the horizon $r=r_h$, while the metric component $g^{rr}$ vanish on the horizon and becomes positive ($g^{rr}>0$) for $r>r_h$. From Eq. (\ref{intpart}) We obtain that
\begin{equation}
\int_{r_h}^{+\infty} dr g^{rr} \sqrt{g}\phi (\partial_r\phi)^{2}=-\int_{r_h}^{+\infty} dr \sqrt{g} \phi V'(\phi)
\end{equation}
As we have assumed that $\phi>0$, from the above equation we conclude that $V'(\phi(r))<0$, at least in a region of the radius $r$ with $r>r_h$.
This confirms the non-convex nature of the potential $V(\phi)$ for 5D scalar black hole solutions which behave asymptotically like an $AdS_5$
Schwarzschild black hole ($n\geq 2$). For $0<n<2$, as we have mentioned previously, the Einstein equation possess black hole solutions (with
a horizon and a singularity, see Fig. \ref{5}) but their asymptotic behavior is not of the form $r^2-\mu/r^2$. Even in this case ($0<n<2$) we have checked numerically that the potential does not possesses a convex form.

\section{Conclusions}

We studied 5D scalar black hole solutions, with flat 3D slices, by using a semi-analytical way. In our approach first we chose the scalar field $\phi(r)$ and subsequently the corresponding self-interacting potential $V(\phi)$ was determined.

We focused to continuous scalar field configurations
which are nonzero in a small region of r near the origin, and vanish rapidly in the large r limit. In particular, two cases for the black hole scalar hair were examined: (a) an exponential decaying scalar field profile $\phi=\phi_0 e^{-r/d}$ and (b) an inverse power scalar field profile $\phi=q/r^n$. The constants of integration in the metric of the scalar black hole solution are fixed by assuming an asymptotic behavior identical with that of $AdS_5$ Schwarzschild black hole. We show that this asymptotic behavior can be achieved only if the scalar field vanish asymptotically like $1/r^2$ or faster. However, we constructed black hole solutions (with a horizon and a singularity for the profile (b)) even in the interval $0<n<2$, but we show that they have not an asymptotic behavior of the form $r^2-\mu/r^2$.

 We found that the reconstructed potential $V(\phi)$, for the cases (a) and (b), has a double well form (see Fig. \ref{3} above) however as
 we see it blows up near the singularity of the black hole $\phi(r_s)$, hence the scalar field is restricted in the region $0\leq \phi \leq \phi(r_s)$. We also demonstrated (and checked numerically) a "no hair theorem" for our case, which requires a non-convex form for the potential $V(\phi)$, or equivalently that $V'(\phi(r))$ is negative at least in a region of $r$ with $r>r_h$, in order to have a nontrivial solution for the scalar field. Although, this theorem has been proven only when the scalar field vanish asymptotically like $1/r^2$ or faster, our numerical results show that the non-convex nature of the potential remains even in the region $0<n<2$ for the profile (b).

 Finally, we would like to state that the stability of our solutions, against linear scalar field perturbations, has not been examined in this work and it is left for further investigation.

\section{Acknowledgements}

We would like to thank Professor G. Koutsoumbas for reading and commenting on the manuscript.


\begin{thebibliography}{99}

\bibitem{RW}
R. ~Ruffini and J.~ A.~ Wheeler, Phys . Today 24(1), 30 (1971).

%\cite{Bekenstein:1971hc}
\bibitem{Bekenstein:1971hc}
  J.~D.~Bekenstein,
  %``Nonexistence of baryon number for static black holes,''
  Phys.\ Rev.\  D {\bf 5} (1972) 1239.
  %%CITATION = PHRVA,D5,1239;%%

%\cite{Bekenstein:1974sf}
\bibitem{Bekenstein:1974sf}
  J.~D.~Bekenstein,
  %``Exact Solutions Of Einstein Conformal Scalar Equations,''
  Annals Phys.\  {\bf 82} (1974) 535.
  %%CITATION = APNYA,82,535;%%

 %\cite{Bechmann:1995sa}
\bibitem{Bechmann:1995sa}
  O.~Bechmann and O.~Lechtenfeld,
  %``Exact black hole solution with selfinteracting scalar field,''
  Class.\ Quant.\ Grav.\  {\bf 12} (1995) 1473
  [arXiv:gr-qc/9502011].
  %%CITATION = CQGRD,12,1473;%%
%\cite{Bronnikov:2001ah}

\bibitem{Dennhardt:1996cz}
  H.~Dennhardt and O.~Lechtenfeld,
  %``Scalar deformations of Schwarzschild holes and their stability,''
  Int.\ J.\ Mod.\ Phys.\  A {\bf 13} (1998) 741
  [arXiv:gr-qc/9612062].
  %%CITATION = IMPAE,A13,741;%%

\bibitem{Bronnikov:2001ah}
  K.~A.~Bronnikov and G.~N.~Shikin,
  %``Spherically symmetric scalar vacuum: No-go theorems, black holes and
  %solitons,''
  Grav.\ Cosmol.\  {\bf 8} (2002) 107
  [arXiv:gr-qc/0109027].
  %%CITATION = GRCOF,8,107;%%

%\cite{Nucamendi:1995ex}
\bibitem{Nucamendi:1995ex}
  U.~Nucamendi and M.~Salgado,
  %``Scalar hairy black holes and solitons in asymptotically flat spacetimes,''
  Phys.\ Rev.\  D {\bf 68} (2003) 044026
  [arXiv:gr-qc/0301062].
  %%CITATION = PHRVA,D68,044026;%%

 %\cite{Hertog:2006rr}
\bibitem{Hertog:2006rr}
  T.~Hertog,
  %``Towards a Novel no-hair Theorem for Black Holes,''
  Phys.\ Rev.\  D {\bf 74} (2006) 084008
  [arXiv:gr-qc/0608075].
  %%CITATION = PHRVA,D74,084008;%%

%\cite{Kuriakose:2008dj}
\bibitem{Kuriakose:2008dj}
  P.~I.~Kuriakose and V.~C.~Kuriakose,
  %``Static Black Hole dressed with a massive Scalar field,''
  arXiv:0805.4554 [gr-qc].
  %%CITATION = ARXIV:0805.4554;%%

 %\cite{Martinez:2004nb}
\bibitem{Martinez:2004nb}
  C.~Martinez, R.~Troncoso and J.~Zanelli,
  %``Exact black hole solution with a minimally coupled scalar field,''
  Phys.\ Rev.\  D {\bf 70}, 084035 (2004)
  [arXiv:hep-th/0406111].
  %%CITATION = PHRVA,D70,084035;%%
%\cite{Dotti:2007cp}
\bibitem{Dotti:2007cp}
  G.~Dotti, R.~J.~Gleiser and C.~Martinez,
  %``Static black hole solutions with a self interacting conformally coupled
  %scalar field,''
  Phys.\ Rev.\  D {\bf 77} (2008) 104035
  [arXiv:0710.1735 [hep-th]].
  %%CITATION = PHRVA,D77,104035;%%
%\cite{Torii:2001pg}
\bibitem{Torii:2001pg}
  T.~Torii, K.~Maeda and M.~Narita,
  %``Scalar hair on the black hole in asymptotically anti-de Sitter spacetime,''
  Phys.\ Rev.\  D {\bf 64} (2001) 044007.
  %%CITATION = PHRVA,D64,044007;%%



%\cite{Winstanley:2002jt}
\bibitem{Winstanley:2002jt}
  E.~Winstanley,
  %``On the existence of conformally coupled scalar field hair for black  holes
  %in (anti-)de Sitter space,''
  Found.\ Phys.\  {\bf 33} (2003) 111
  [arXiv:gr-qc/0205092].
  %%CITATION = FNDPA,33,111;%%

%\cite{Winstanley:2005fu}
\bibitem{Winstanley:2005fu}
  E.~Winstanley,
  %``Dressing a black hole with non-minimally coupled scalar field hair,''
  Class.\ Quant.\ Grav.\  {\bf 22}, 2233 (2005)
  [arXiv:gr-qc/0501096].
  %%CITATION = CQGRD,22,2233;%%
  
  %\cite{Zeng:2009fp}
\bibitem{Zeng:2009fp}
  D.~f.~Zeng,
  An Exact Hairy Black Hole Solution for AdS/CFT Superconductors,
  arXiv:0903.2620 [hep-th].
  %%CITATION = ARXIV:0903.2620;%%


  
%\cite{Harper:2003wt}
\bibitem{Harper:2003wt}
  T.~J.~T.~Harper, P.~A.~Thomas, E.~Winstanley and P.~M.~Young,
  %``Instability of a four-dimensional de Sitter black hole with a  conformally
  %coupled scalar field,''
  Phys.\ Rev.\  D {\bf 70} (2004) 064023
  [arXiv:gr-qc/0312104].
  %%CITATION = PHRVA,D70,064023;%%

%\cite{Hertog:2004bb}
\bibitem{Hertog:2004bb}
  T.~Hertog and K.~Maeda,
  %``Stability and thermodynamics of AdS black holes with scalar hair,''
  Phys.\ Rev.\  D {\bf 71} (2005) 024001
  [arXiv:hep-th/0409314].
  %%CITATION = PHRVA,D71,024001;%%

%\cite{Mavromatos:1995fc}
\bibitem{Mavromatos:1995fc}
  N.~E.~Mavromatos,
  %``Eluding the no-hair conjecture for black holes,''
  arXiv:gr-qc/9606008.
  %%CITATION = GR-QC/9606008;%%

%\cite{Weinberg:2001gc}
\bibitem{Weinberg:2001gc}
  E.~J.~Weinberg,
  %``Black holes with hair,''
  arXiv:gr-qc/0106030.
  %%CITATION = GR-QC/0106030;%%

\bibitem{antoniadis} I.~Antoniadis,
  %``A Possible new dimension at a few TeV,''
  Phys.\ Lett.\  B {\bf 246} (1990) 377.
  %%CITATION = PHLTA,B246,377;%%

%\cite{ArkaniHamed:1998rs}
\bibitem{ArkaniHamed:1998rs}
  N.~Arkani-Hamed, S.~Dimopoulos and G.~R.~Dvali,
  %``The hierarchy problem and new dimensions at a millimeter,''
  Phys.\ Lett.\  B {\bf 429} (1998) 263
  [arXiv:hep-ph/9803315].
  %%CITATION = PHLTA,B429,263;%%

%\cite{Antoniadis:1998ig}
\bibitem{Antoniadis:1998ig}
  I.~Antoniadis, N.~Arkani-Hamed, S.~Dimopoulos and G.~R.~Dvali,
  %``New dimensions at a millimeter to a Fermi and superstrings at a TeV,''
  Phys.\ Lett.\  B {\bf 436}, 257 (1998)
  [arXiv:hep-ph/9804398].
  %%CITATION = PHLTA,B436,257;%%

 %\cite{Randall:1999ee}
\bibitem{Randall:1999ee}
  L.~Randall and R.~Sundrum,
  %``A large mass hierarchy from a small extra dimension,''
  Phys.\ Rev.\ Lett.\  {\bf 83} (1999) 3370
  [arXiv:hep-ph/9905221].
  %%CITATION = PRLTA,83,3370;%%

%\cite{Randall:1999vf}
\bibitem{Randall:1999vf}
  L.~Randall and R.~Sundrum,
  %``An alternative to compactification,''
  Phys.\ Rev.\ Lett.\  {\bf 83} (1999) 4690
  [arXiv:hep-th/9906064].
  %%CITATION = PRLTA,83,4690;%%


%\cite{Csaki:2000dm}
\bibitem{Csaki:2000dm}
  C.~Csaki, J.~Erlich and C.~Grojean,
  %``Gravitational Lorentz violations and adjustment of the cosmological
  %constant in asymmetrically warped spacetimes,''
  Nucl.\ Phys.\  B {\bf 604} (2001) 312
  [arXiv:hep-th/0012143].
  %%CITATION = NUPHA,B604,312;%%
  %\cite{Farakos:2008rv}

\bibitem{Farakos:2008rv}
  K.~Farakos, N.~E.~Mavromatos and P.~Pasipoularides,
  %``Bulk photons in Asymmetrically Warped Space-times and Non-trivial Vacuum
  %Refractive Index,''
  JHEP {\bf 0901} (2009) 057
  [arXiv:0807.0870 [hep-th]];
  %%CITATION = JHEPA,0901,057;%%
%\cite{Farakos:2009ka}
  K.~Farakos, N.~E.~Mavromatos and P.~Pasipoularides,
  %``Asymmetrically Warped Brane Models, Bulk Photons and Lorentz Invariance,''
  arXiv:0902.1243 [hep-th];
  %%CITATION = ARXIV:0902.1243;%%
%\cite{Farakos:2009ui}
  K.~Farakos,
  Lorentz violation effects in asymmetric two brane models: a nonperturbative
  analysis,
  arXiv:0903.3356 [hep-th].
  %%CITATION = ARXIV:0903.3356;%%

%\cite{Gubser:1999vj}
\bibitem{Gubser:1999vj}
  S.~S.~Gubser,
  %``AdS/CFT and gravity,''
  Phys.\ Rev.\  D {\bf 63} (2001) 084017
  [arXiv:hep-th/9912001].
  %%CITATION = PHRVA,D63,084017;%%




\end{thebibliography}
\end{document}